# Direct *ab initio* simulation of silver ion dynamics in chalcogenide glasses

**De Nyago Tafen[1], D. A. Drabold[*1] and M. Mitkova[2]**

[1] Department of Physics and Astronomy, Ohio University, Athens OH 45701, USA
[2] Center for Solid State Electronics Research, Arizona State University, Tempe, AZ 85287-6206, USA

In this paper, we present new models of germanium selenide chalcogenide glasses heavily doped with silver. The models were readily obtained with *ab initio* molecular dynamics and their structure agrees closely with diffraction measurements. Thermal molecular dynamics simulation reveals the dynamics of $Ag^+$ ions and the existence of trapping centers as conjectured in other theory work. We show that first principles simulation is a powerful tool to reveal the motion of ions in glass.

Mobile ions in amorphous materials have been a serious object of investigation [1], and their dynamics in disordered hosts constitute one of the major unsolved problems in the field of solid state ionics. Certain compositions of silver doped GeSe glasses (close to compositions reported here) have practical potential for novel computer memory devices [2]. The structure of the Ge-Se-Ag glasses has been investigated using standard techniques, including X-ray diffraction [4, 5] neutron diffraction with isotopic substitution [6], EXAFS [7], differential anomalous X-ray scattering (DAS) [8, 9], Modulated Differential Scanning Calorimetry (MDSC) and Raman spectroscopy [10]. Despite this database, the structure of the ternary Ge-Se-Ag glasses has not yet been completely determined. There continues to be a debate on basic aspects of the glass structure (i.e. homogeneity and Ag coordination) especially for Se rich glasses with more than 67 % Se. In this work, we focused on Ag-doped glasses containing Ge 25 at. % and Se 75 at. % that we label later as $GeSe_3$. Such compositions are of special interest because of proximity to the intermediate phase, which is reported to not age and thus might contribute to formation of memory or other type of devices with special reliability [10].

For the simulations reported in this paper, we use FIREBALL2000 developed by Lewis and co-workers [11]. Total energies and forces were computed within an *ab initio* local orbital formalism [12]. The exchange-correlation energy was treated within the LDA, for which we used the results of Ceperley and Alder [13], as interpolated by Perdew and Zunger [14]. The pseudopotential and pseudoatomic wave functions were generated in the Troullier-Martins form [15] employing the scheme of Fuchs and Scheffler [16].

The models described here were generated using the melt-quenching method. We placed atoms randomly in a cubic supercell according to the correct stoichiometry [for $(GeSe_3)_{0.90}Ag_{0.10}$ 54 germanium atoms, 162 selenium atoms and 24 silver atoms; for $(GeS_3)_{0.85}Ag_{0.15}$ 51 germanium atoms, 153 selenium atoms and 36 silver atoms] with the minimum distance between atoms 2Å. The size of the cubic cells was chosen to make the density of these glasses match experimental data. The box size of the 240 atom supercell of $(GeSe_3)_{0.90}Ag_{0.10}$ and $(GeSe_3)_{0.85}Ag_{0.15}$ are respectively 18.601 Å and 18.656 Å with corresponding density 4.98 g/cm$^3$ and 5.03 g/cm$^3$ [5]. The cells were annealed, and we obtained well thermalized melts at 4800 K. We took three steps to cool the cells. First, they were equilibrated at 1100 K for 3 ps; then they were slowly cooled to 300 K for approximately 5 ps. In the final step, the cells were steepest descent quenched to 0 K and all forces were smaller in magnitude than 0.02 eV/ Å. Additional details will be reported elsewhere.

The structure of these models is analyzed by computing the structure factor. Fig. 1 shows the calculated static structure factors for $(GeSe_3)_{0.90}Ag_{0.10}$ and $(GeSe_3)_{0.85}Ag_{0.15}$ and the comparison with the experimental data from Ref. [5]. Our calculation reveals gratifying agreement with experiment.

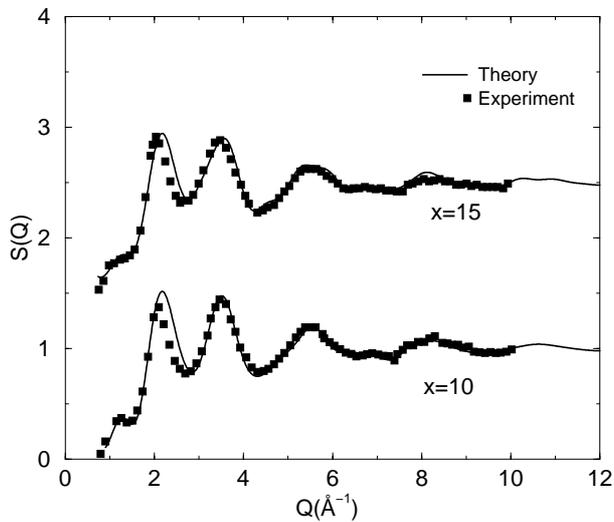

**Figure 1** Calculated total structure factor S(Q) of (GeSe$_3$)$_{0.90}$Ag$_{0.10}$ ($x = 10$) and (GeSe$_3$)$_{0.85}$Ag$_{0.15}$ ($x = 15$) glasses compared to experimental data of ref. [5]

To explore the mechanism of diffusion of silver, we examined the trajectories of these particles. We obtained $2.5\times10^4$ time steps of time development, for a total time of 62.5 ps and a fixed temperature of 1000K. Fig. 2 illustrates 2D projections of trajectories of the most and least mobile Ag$^+$ ions in each model. For short times, the mean-squared displacement of the most mobile atoms increases due to the diffusive motion of Ag. At intermediate times, the atoms may be trapped in a cage formed by their neighbors, and at the longest times we can explore, they can escape such traps, diffuse and again be trapped. Thus, our trajectories can largely be separated into vibration around stable sites and hops between such sites. This is an explicit confirmation of the existence of Scher-Lax-Phillips traps, discussed in the context of relaxation processes in disordered systems[17]. In both glasses, a significant fraction of Ag atoms do move large distances. In (GeS$_3$)$_{0.90}$Ag$_{0.10}$, about 62.5 % of silver atoms move an average distance greater than 2.5 Å for a time scale of 20 ps. Among them 46 % have an average displacement greater than 3 Å. By contrast only 17 % of Ag atoms move less than 2 Å. On the other hand, about 70 % of Ag atoms move on an average distance greater than 2.5 Å in (GeS$_3$)$_{0.85}$Ag$_{0.15}$ for a period of 25 ps. 11.11 % of those atoms have an average displacement greater than 5 Å. The most mobile Ag atoms move an average distance of 7 Å in this time. These numbers illustrate the high ionic mobility of Ag ions in these complex glasses.

Further insight into the mechanism of diffusion in (Ge$_x$Se$_{1-x}$)$_{1-y}$Ag$_y$ can be found by studying the behavior of the local atomic density of particular regions containing silver atoms. Hence we calculate the local density of the most and least mobile silver atoms as a function of time, then compare them to the density of the glass. To do so, we construct a sphere of radius R=4 Å. The center of the sphere is the position of the Ag atoms we are tracking at a time t (the sphere moves with the atom). Then we calculate the average density in the spherical volume. Fig. 3 illustrates the local density of Ag atoms as a function of time. As seen on the figure, the more mobile Ag atoms are consistently located in regions with a lower local density (lower local volume fraction) and higher disorder. This argues for a simple ``free volume picture" of Ag diffusion.

This work highlights the existence of trapping centers[17] and explicitly illustrates the trapping and release processes from thermal MD simulation. Detailed discussion will be presented elsewhere [18]. In the longer term, statistics of trap lifetimes and hopping rates will be obtained. Ideally, it might prove possible to compute the inputs into phenomenological hopping/trapping models from first principles dynamical simulation.

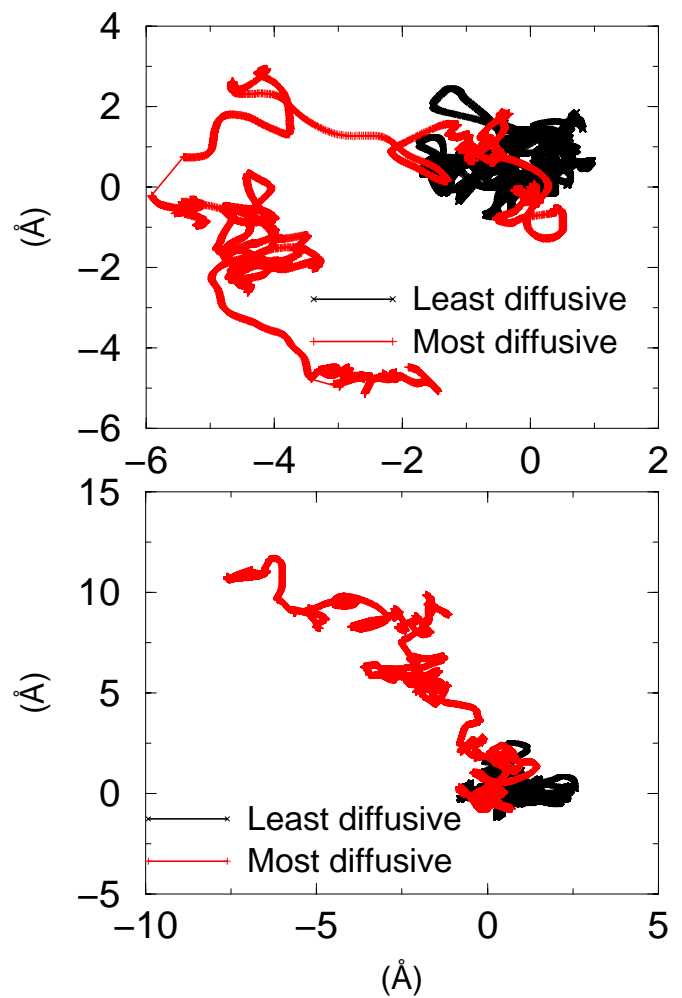

**Figure 2** Trajectories of the most and least diffusive Ag atoms in (GeSe$_3$)$_{0.90}$Ag$_{0.10}$ (top panel) and (GeSe$_3$)$_{0.85}$Ag$_{0.15}$ (bottom panel) glasses (T=1000 K).

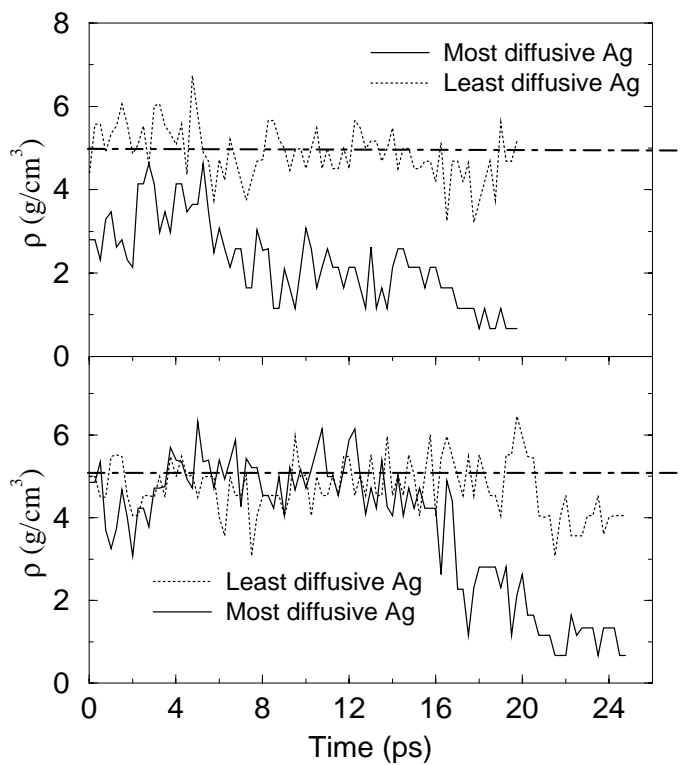

**Figure 3** Local density in vicinity of the slowest and fastest diffusing Ag$^+$ ions as a function of time compared to the mean density (horizontal dot-dashed line) in (GeSe$_3$)$_{0.90}$Ag$_{0.10}$ (top panel) and (GeSe$_3$)$_{0.85}$Ag$_{0.15}$ (bottom panel) glasses.

We thank the US NSF for support under grants DMR-0074624, DMR-0205858 and DMR-0310933. We also gratefully acknowledge the support of Axon Technologies, Inc. We thank Prof. Michael Kozicki and Dr. J. C. Phillips for helpful conversations and support.

[1] C.A. Angell, Ann. Rev. Phys. Chem. **43**, 693 (1992).
[2] N. M. Mitkova and M.N Kozicki, J. Non-Cryst. Sol. **299-302**, 1023 (2002).
[3] A. Pradel, N. Kuvata, M. Ribes, J. Phys.: Condens. Matter **15**, S1561-S1571 (2003).
[4] Fischer-Colbrie, A. Bienenstock, P.H. Fuoss, M.A. Marcus, Phys. Rev. B **38**, 12388 (1988).
[5] A. Piarristeguy, M. Fontana and B. Arcondo, J. Non-Cryst. Sol. **332**, 1 (2003).
[6] J.H. Lee, A.P. Owens, S.R. Elliott, J. Non-Cryst. Sol. **164-166**, 139 (1993).
[7] J.M. Oldale, J. Rennie, and S.R. Elliott, Thin Solid Films **164**, 467 (1988).
[8] R. J. Dejus, D.J. Le Poire, S. Susman, K. J. Volin, and D. L. Price, Phys. Rev. B **44**, 11705 (1991); R. J. Dejus, S. Susman, K. J. Volin, D. G Montague, and D. L Price, J. Non-Cryst. Sol. **143**, 162 (1992).
[9] J.D Westwood, P. Georgopoulos, and D.H. Whitmore, J. Non-Cryst. Sol. **107**, 88 (1988).
[10] M. Mitkova, Yu Wang, and P. Boolchand, Phys. Rev. Lett. **19**, 3848 (1999); Y. Wang, M. Mitkova, D.G. Georgiev, S. Mamedov, and P. Boolchand, J.Phys.: Condens. Matter **15**, S1573-S1584 (2003).
[11] J. P. Lewis, K. R. Glaesemann, G. A. Voth, J. Fritsch, A.A. Demkov, J. Ortega, and O.F. Sankey, Phys. Rev. B **64**, 195103 (2001).
[12] O. Sankey and D. J. Nikleswki, Phys. Rev. B **40**, 3979 (1989).
[13] D. M. Ceperley and B. J. Alder, Phys. Rev. Lett. **45**, 566 (1980).
[14] J. P. Perdew and A. Zunger, Phys. Rev. B **23**, 5048 (1981).
[15] N. Troullier and J. L. Martins, Phys. Rev. B **43**, 1993 (1991).
[16] M. Fuchs and M. Scheffler, Comput. Phys. Commun. **119**, 67 (1999).
[17] J. C. Phillips, Rep. Prog. Phys **59** 1133 (1996).
[18] D. N. Tafen, D. A. Drabold and M. Mitkova (unpublished).